\documentclass{svjour3}                     
\smartqed  
\usepackage{graphicx}
\journalname{Few-Body Systems}
\begin{document}


\title{Relativistic descriptions of few-body systems\footnote{Presented at the
21st European Conference on Few-Body Problems in Physics,
Salamanca, Spain, 30 August - 3 September 2010.}}


\titlerunning{Relativistic descriptions of few-body systems}

\author{V.A.~Karmanov        
}

\authorrunning{V.A.~Karmanov} 

\institute{V.A.~Karmanov \at Lebedev
Physical Institute, Leninsky prospekt 53, 119991 Moscow, Russia \\
              \email{karmanov@sci.lebedev.ru}           
}

\date{Received: date / Accepted: date}

\maketitle

\begin{abstract}
A brief review of relativistic effects in few-body systems, of
theoretical approaches, recent developments and applications is
given.  Manifestations of relativistic effects in the binding
energies, in the electromagnetic form factors and in three-body
observables are demonstrated. The three-body forces of
relativistic origin are also discussed.
\end{abstract}

\section{Introduction}
\label{intro} Light nuclei give a typical example of a few-body
weakly bound system. Their binding energies $B$ are of the order
of 0.1\% from their masses $M$. This however does not mean that
the relativistic effects in light nuclei are also so small: they
are much larger than the ratio $B/M$. The reason is that, in
contrast to the hydrogen atom (for instance), the small nuclear
binding energy is a results of cancellation of much more
significant kinetic and potential energies. This can be
illustrated as follows.

Let us consider a system of two non-relativistic particles
interacting by the Yukawa potential:
\mbox{$V_{non-rel.}(r)=-\frac{\alpha}{r}\exp(-\mu r)$} that in the
momentum space corresponds to the kernel:
\begin{equation}\label{Vnr}
V_{non-rel.}(\vec{q})= \frac{-4\pi\alpha}{\mu^2+\vec{q}^2}.
\end{equation}
This non-relativistic kernel is a limiting case of the following
relativistic one-boson exchange kernel:
\begin{equation}\label{Vr}
V_{rel.}(q)= \frac{-4\pi\alpha}{\mu^2-q^2-i\epsilon} =
\frac{-4\pi\alpha}{\mu^2+\vec{q}^2-q_0^2-i\epsilon}.
\end{equation}
$V_{rel.}$ turns into $V_{non-rel.}$ when $q_0\to 0$. The kernel
$V_{non-rel.}$ enters the Shr\"odinger equation. The kernel
$V_{rel.}$ enters the relativistic Bethe-Salpeter (BS) equation
\cite{BS}. Another popular relativistic approach is light-front
dynamics (LFD). The corresponding equation -- light front (LF)
equation, see for review \cite{cdkm,bpp}, -- contains the kernel
$V_{rel.}$, which analytical form differs from (\ref{Vr}), but its
non-relativistic limit also coincides with (\ref{Vnr}).
\begin{figure}
\begin{center}
\includegraphics[width=5.5cm]{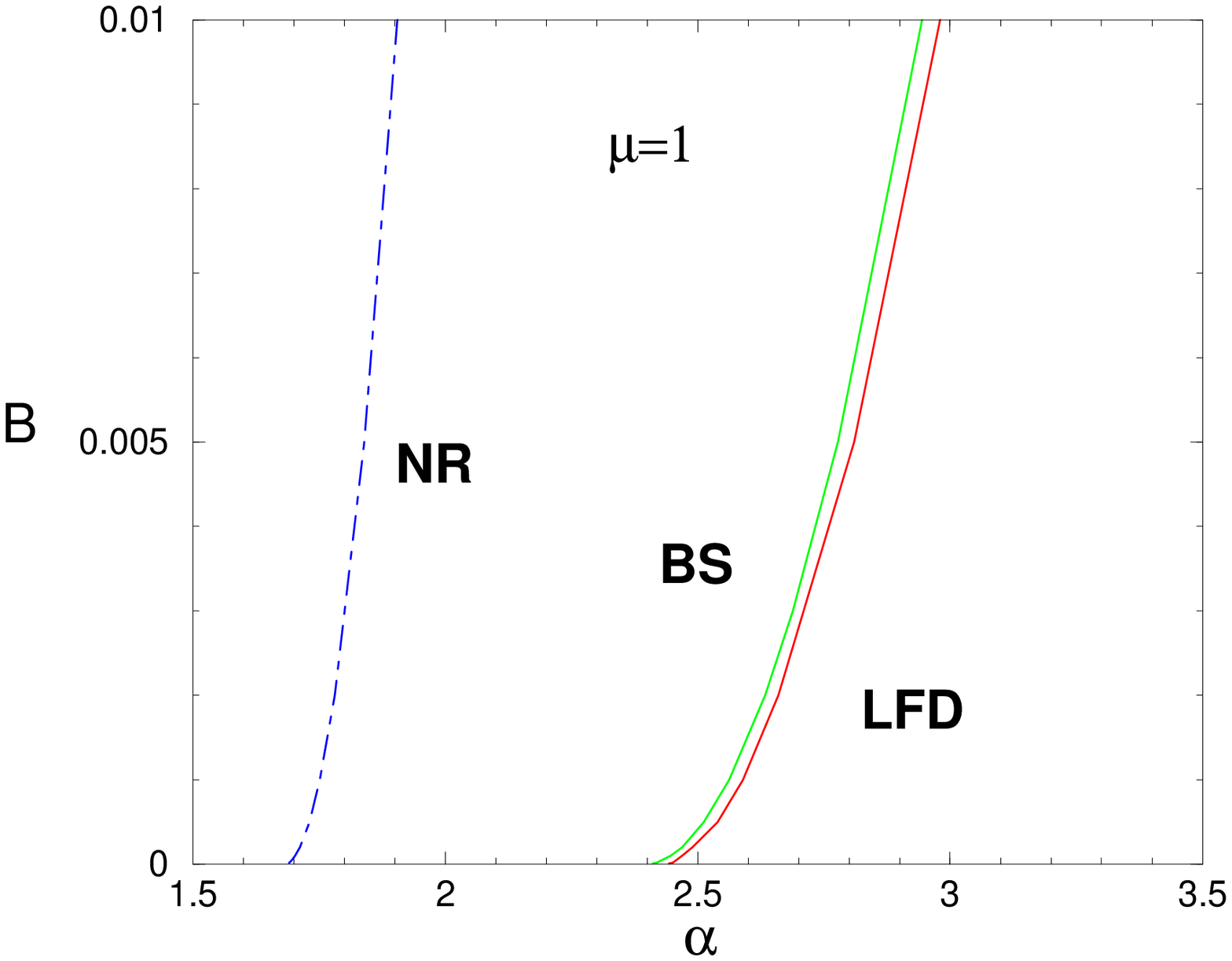}
\hspace{0.5cm}
\includegraphics[width=5.5cm]{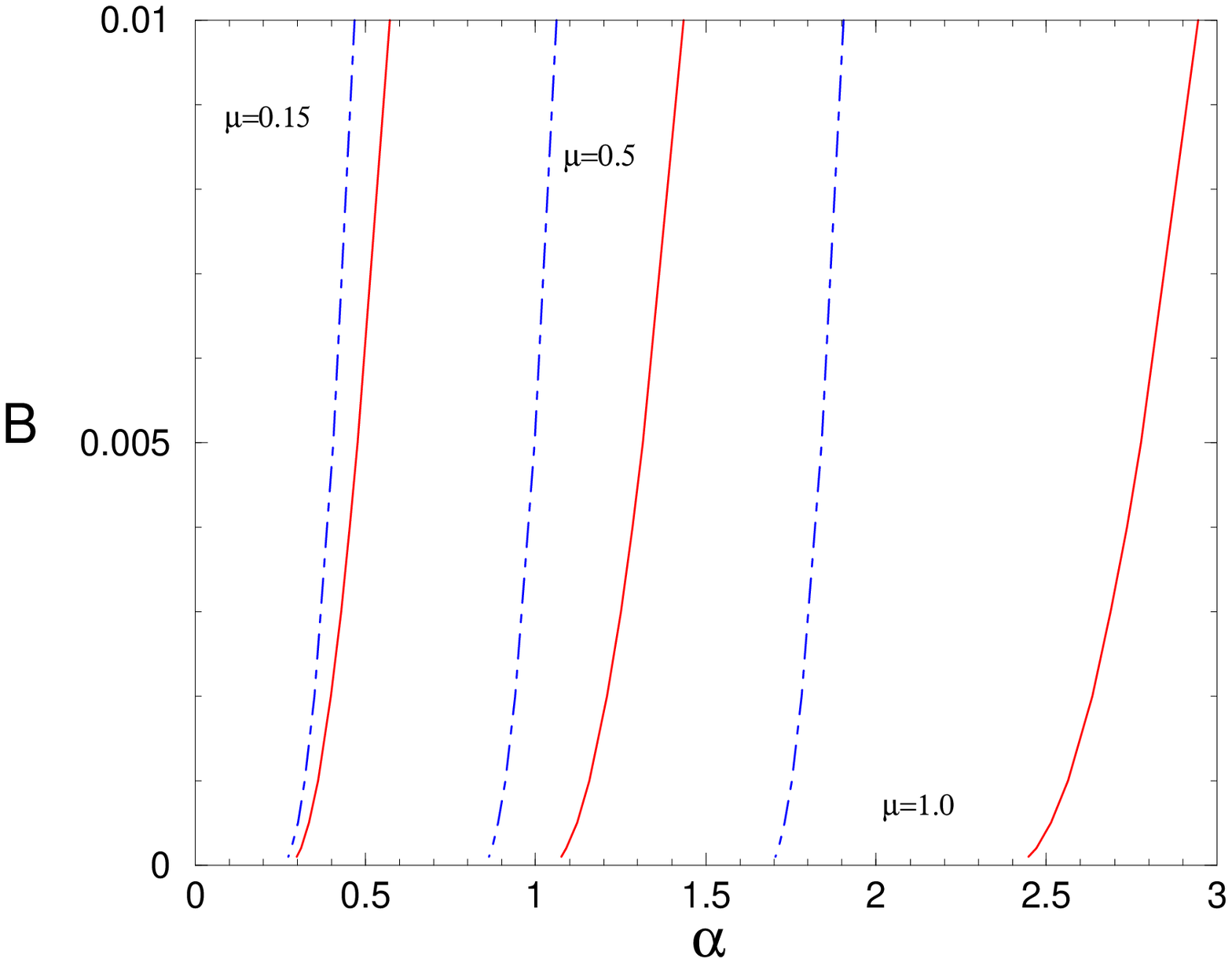}
\caption{Left: Binding energy $B$ vs. coupling constant $\alpha$,
found via Shr\"odinger equation (NR), Bethe-Salpeter (BS) and the
light-front (LFD) one, for exchange mass $\mu=1$ in the kernels
(\ref{Vnr}) and (\ref{Vr}). Right: the same as at left, but for
different exchange masses $\mu$ (here the BS and LFD results are
indistinguishable from each other). The results are from
\cite{MC_2000}.}
\label{fig1}       
\end{center}
\end{figure}
All these three equations -- Shr\"odinger equation, BS and LF ones
-- were solved, for spinless particles in the $J=0$ state, in Ref.
\cite{MC_2000}. The results in the limit of extremely small
binding energy $B\to 0$ are shown in fig. \ref{fig1}. Left panel
corresponds to heavy exchange mass $\mu=1$ (in the units of the
constituent mass $m=1$). We see that the relativistic calculations
BS and LF are very close to each other and, at the same time, they
strongly differ from the non-relativistic result NR. This shows
that the relativistic effects can be important even at small
binding energy. The curves at the right panel demonstrate that
when the exchange mass decreases, the difference between
relativistic and non-relativistic results decreases too. That is,
the true relativistic system is a system not only with very small
binding energy but also with interaction resulted from exchange by
zero mass. When the exchange mass is not small (of the order of
the constituent mass $m$), then the particles are in a very narrow
potential well of the radius $r\sim 1/m$. Then their momenta are
comparable with their masses $k\sim 1/r\sim m$, the kinetic energy
(and the system at all) is relativistic  and, hence, the small
binding energy is a result of cancellation of large (positive)
kinetic and large (negative) potential energies. That's why a
system with small binding energy may be still relativistic.
Similar situation is realized in nuclei, since the exchange mesons
like $\omega$ and $\rho$ are enough heavy ($\mu_{\omega}\approx
\mu_{\rho}\approx 0.8\,m$).

In the opposite case of strongly bound systems, there is an
example \cite{dshvk} that even a system with extremely large
binding energy, such that its total mass tends to zero, may be
dominated (by 90\%) by a few lowest Fock sectors, containing two,
three and four particles. It still remains to be a {\em few-body}
system (though, highly relativistic).

Even in the case of mainly non-relativistic system (average
momentum is very small), its impulse distribution contains a
relativistic tail. This tail may be very small, but it completely
determines the e.m. form factor at large momentum transfer. The
form factor of this system is also very small but it should be
calculated in a relativistic approach.

All that requires development of appropriate approaches to
relativistic description of few-body systems. Brief review of
these approaches is given in the next section.

\section{Relativistic descriptions}
\label{sec1} In non-relativistic quantum mechanics the wave
function is an eigenvector of Hamiltonian: $H\psi=E\psi$. Dynamics
is introduced by the adding to free Hamiltonian $H^{(0)}$ an
interaction term $H^{int}=V$:
$$H^{(0)}\to H=H^{(0)}+H^{int}=\frac{p^2}{2m}+V(r).
$$

In relativistic case, the relativistic covariance is guaranteed if
the wave function (or the state vector $|p\rangle$) is forming a
representation of the Poincar\'e group. The latter is determined
by ten generators $P_{\mu},J_{\mu\nu}$ which satisfy the following
commutation relations:
\begin{eqnarray}\label{comrel}
&&[P_{\mu},P_{\nu}]=0, \quad 
[P_{\mu},J_{\kappa\rho}]
=i(g_{\mu\rho}P_{\kappa} -g_{\mu\kappa}P_{\rho}),
\nonumber\\
&&[J_{\mu\nu},J_{\rho\gamma}]=i(g_{\mu\rho} J_{\nu\gamma}
-g_{\nu\rho}J_{\mu\gamma} +g_{\nu\gamma}J_{\mu\rho}
-g_{\mu\gamma}J_{\nu\rho}).
\end{eqnarray}

The Hamiltonian $H=P_0$ is now only one of generators. Similarly
to non-relativistic case, dynamics is introduced by the adding to
free Poincar\'e generators an interaction terms in a way which
keeps the commutation relations (\ref{comrel}) unchanged. This is
not simple but a solvable task. It can be realized in the
framework of two different approaches: ({\it i}) relativistic
quantum mechanics with fixed number of particles and ({\it ii})
field theory.

In relativistic quantum mechanics the Poincar\'e generators are
the functions of fixed number (say, two or three) of the particles
momenta. The interaction is a phenomenological one, it is fitted
to describe e.g. the two-body phase shifts. Then one can make
predictions: to calculate for instance the e.m. form factors or
three-body observables. For good reviews of this approach see
\cite{Coester,Polyzou}.

In field theory the  Poincar\'e  generators are derived from
Lagrangian by a well-known formulas, given almost in any textbook.
If Lagrangian is not free (contains interaction), then the
interaction appears also in the generators. The state vector
$|p\rangle$, on which the generators act, can be decomposed in the
basis of free fields (similarly to the Fourier decomposition in
plane waves of the non-relativistic wave function). This basis is
represented as an (infinite) set of Fock components with
increasing numbers of particles. In practice, this decomposition
is truncated (the desired number of particles is fixed by hand).
After that the approach becomes approximate.

So, in practice, two approaches -- ({\it i}) and ({\it ii}) --
differ from each other by the point where one makes this
truncation: ({\it i}) either from the very beginning, with further
phenomenological construction the generators; ({\it ii}) or after
finding the generators by the field theory recepees. In the latter
case, the kernel is motivated by a field-theoretical Lagrangian.
In its turn, this field-theoretical interaction is mainly
restricted by the one-boson-exchange.

One should also distinguish three forms of relativistic dynamics,
proposed by Dirac \cite{Dirac49}, which exist in both approaches.
Namely: (a) instant form; (b) front form; (c) point form. They
differ from each other by the ways of introducing the interaction
in generators. In the instant form the time component $P_0$ of the
four-momentum operator $P_{\mu}$ contains interaction, whereas the
spatial components $P_j$ ($j=x,y,z$) are free. The interaction
enters also in the components $J_{0j}$ of the operators
$J_{\mu\nu}$. The components $J_{ij}$ are free. In the front form
(LFD) the interaction enters the component $P_-=P_0-P_z$, whereas
the components $P_+=P_0+P_z,\;P_x,\;P_y$ are free. The operator
$J_{\mu\nu}$ is constructed correspondingly. The components of
this operator which transform the LF plane $t+z=const$ into itself
are free. In the point form all the components of $P_{\mu}$
contain interaction, whereas the operator $J_{\mu\nu}$ is free.

In its turn, LFD is developed in two forms: ordinary LFD with the
LF plane $t+z=const$ (see for review \cite{bpp}) and explicitly
covariant LFD \cite{cdkm} with the LF plane given by the invariant
equation $\omega\cdot x=\omega_0 t-\vec{\omega}\cdot \vec{x}
=const$, where $\omega=(\omega_0,\vec{\omega})$ is a four-vector
with $\omega^2=0$. The main advantage of this latter formulation
is in the fact that the dependence of the state vector on the LF
orientation is given explicitly, in terms of the four-vector
$\omega$ (see e.g. the LF deuteron wave function (\ref{psid})
below). In the particular case $\omega=(1,0,0,-1)$ we come back to
the ordinary formulation of LFD.

The operator of e.m. current $j_{\mu}$ used to calculate e.m. form
factors, like any four-vector operator, has the following
commutation relation with $J_{\kappa\rho}$:
\begin{equation}\label{jJ}
[j_{\mu},J_{\kappa\rho}]= i(g_{\mu\rho}j_{\kappa}
-g_{\mu\kappa}j_{\rho}).
\end{equation}
If $J_{\kappa\rho}$ contains interaction (in l.h.-side of
(\ref{jJ})), then r.h.-side of (\ref{jJ}), i.e.  $j_{\mu}$, also
must contain interaction. This means that in interacting system
the exact e.m. current cannot be free (except for the point form
of dynamics).

Another series of (field-theoretical) relativistic approached
deals not with the state vector $|p\rangle$ itself, but it is
based on the BS amplitude \cite{BS} defined as:
\begin{equation}\label{BS}
\Phi(x_1,x_2,p)= \langle 0|T\left(\varphi(x_1)
\varphi(x_2)\right)|p\rangle,
\end{equation}
$\varphi(x)$ is the Heisenberg field operator and $\langle 0|$ is
the vacuum state. In the momentum space:
$$
\Phi=\Phi(k_1,k_2,p)=\Phi(k,p),\quad p^2=M^2,\;k_1^2\neq
m^2,\;k_2^2\neq m^2 \quad \mbox{and}\quad k=(k_1-k_2)/2.
$$
The BS equation for $\Phi$ is singular, that complicates its
numerical solution. To avoid singularities, one can transform this
equation in the Euclidean space. Corresponding solution provides
the binding energies. However, to calculate e.m. form factors, we
should know the BS amplitude in Minkowski space. The Wick rotation
in terms of the relative momentum $k$ -- the argument of the BS
amplitude $\Phi(k,p)$ -- is not valid in the form factor integral
over $k$.

The methods to solve BS equation in Minkowski space were recently
developed first for the spinless particles  \cite{bs1} and then
for two fermions \cite{2f}. They are based on the Nakanishi
integral representation \cite{nak63}:
\begin{equation}\label{bsint}
\Phi(k,p)=\int_{-1}^1\mbox{d}z'\int_0^{\infty}\mbox{d}\gamma'
\frac{g(\gamma',z')}{\left[  k^2+p\cdot k\; z' +\frac{1}{4}M^2-m^2
-  \gamma' + i\epsilon\right]^3}.
\end{equation}
This integral determines a singular BS amplitude $\Phi(k,p)$.
However, the Nakanishi weight function $g(\gamma,z)$ is not
singular. Substituting BS amplitude in the form (\ref{bsint}) in
the BS equation, one can derive and solve numerically equation for
$g(\gamma,z)$. Then again using (\ref{bsint}) with known
$g(\gamma,z)$, one can express the observables, like e.m. form
factors, through $g(\gamma,z)$ analytically and then compute them
numerically. Another method to solve the BS in Minkowski space is
based on the separable approximation of the kernel (see
\cite{bbpr} and references therein).

There are also a few reductions of the four-dimensional BS
amplitude to a three-dimensional form (still in the Minkowski
space). In this direction, the approach proposed in \cite{Gross}
(covariant spectator theory) is most advanced and well developed.
In the covariant spectator theory, the NN potential was fitted and
applied to the deuteron and three-body problems as well as to the
e.m. form factors \cite{GG,stadler}.

The theoretical activity in studying the relativistic few-body
systems flourishes in all the forms of dynamics and in all the
approaches listed above.

\section{Some applications}
It is clearly demonstrated (see e.g. \cite{GG}) that the
non-relativistic calculations of the $ed$ elastic cross section do
not describe the data at $Q^2\geq 1\;GeV^2/c^2$. One needs to
perform the calculations with true relativistic deuteron wave
function. In the non-relativistic case the latter is determined by
two spin components: S- and D-waves. In relativistic approaches
the number of components depends on approach. In the spectator
theory \cite{Gross}, there are four components. There are six
components in covariant LFD \cite{karm80}. The deuteron BS
amplitude is determined by eight components (see e.g.
\cite{cdkm}).

As an example, we mention the calculation carried out in the
framework of explicitly covariant LFD. In this approach, the
relativistic deuteron wave function has the form \cite{karm80}:
\begin{eqnarray}\label{psid}
&&\vec{\psi}(\vec{k},\vec{n}) = f_1 \frac{1}{\sqrt{2}}\vec{\sigma}
+ f_2
\frac{1}{2}(\frac{3\vec{k}(\vec{k}\cdot\vec{\sigma})}{\vec{k}^2}
-\vec{\sigma})  +
f_3\frac{1}{2}(3\vec{n}(\vec{n}\cdot\vec{\sigma})
-\vec{\sigma}) \\
 && +  f_4\frac{1}{2k}(3\vec{k}(\vec{n}\cdot\vec{\sigma})+
3\vec{n}(\vec{k}\cdot\vec{\sigma}) -
2(\vec{k}\cdot\vec{n})\vec{\sigma}) +
f_5\sqrt{\frac{3}{2}}\frac{i}{k}[\vec{k}\times \vec{n}] +
f_6\frac{\sqrt{3}}{2k}[[\vec{k}\times
\vec{n}]\times\vec{\sigma}],\nonumber
\end{eqnarray}
where $\vec{n}=\vec{\omega}/|\vec{\omega}|$ and $\vec{\sigma}$ are
the Pauli matrices. The vector $\vec{n}$ just provides the
explicit dependence of this wave function on the LF orientation.
The six components $f_{1-6}$ were calculated in \cite{ck95}. Only
three of them dominate: $f_1,f_2$ (which turn into the S- and
D-waves in non-relativistic limit) and $f_5$, whereas
$f_3,f_4,f_6$ are negligible. The corresponding deuteron e.m. form
factors were calculated  in \cite{ck99}. The results of this
calculation are in good coincidence with the appeared later
experimental data \cite{abbott}. We do not give this comparison
here. A detailed review can be found in \cite{GG}. One can
conclude that the relativistic effects in a two-body system are
taken into account satisfactory.

\begin{figure}
\centering
\includegraphics[width=5.5cm,height=5cm]{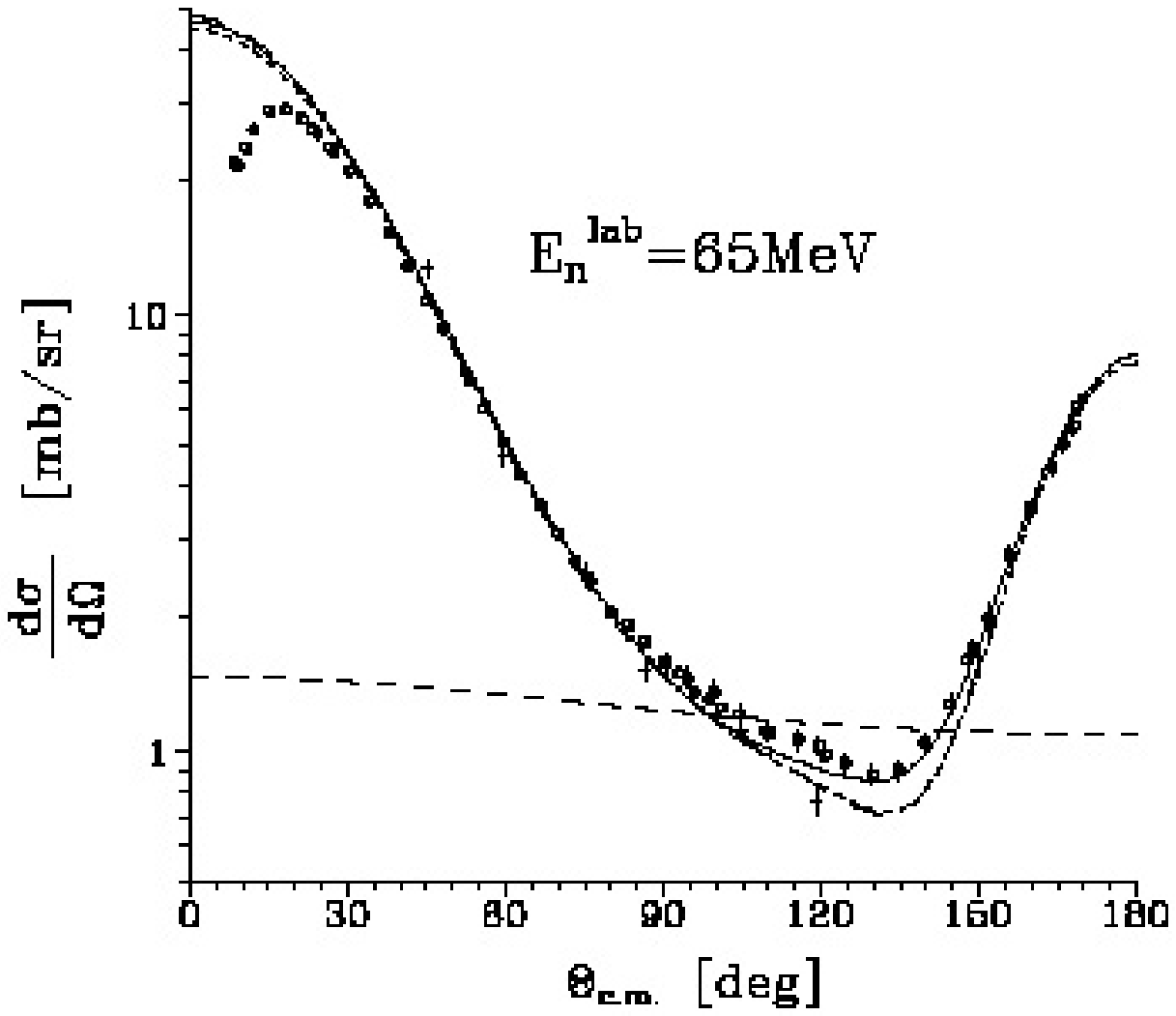}
\hspace{0.5cm}
\includegraphics[width=5.5cm,height=5.5cm]{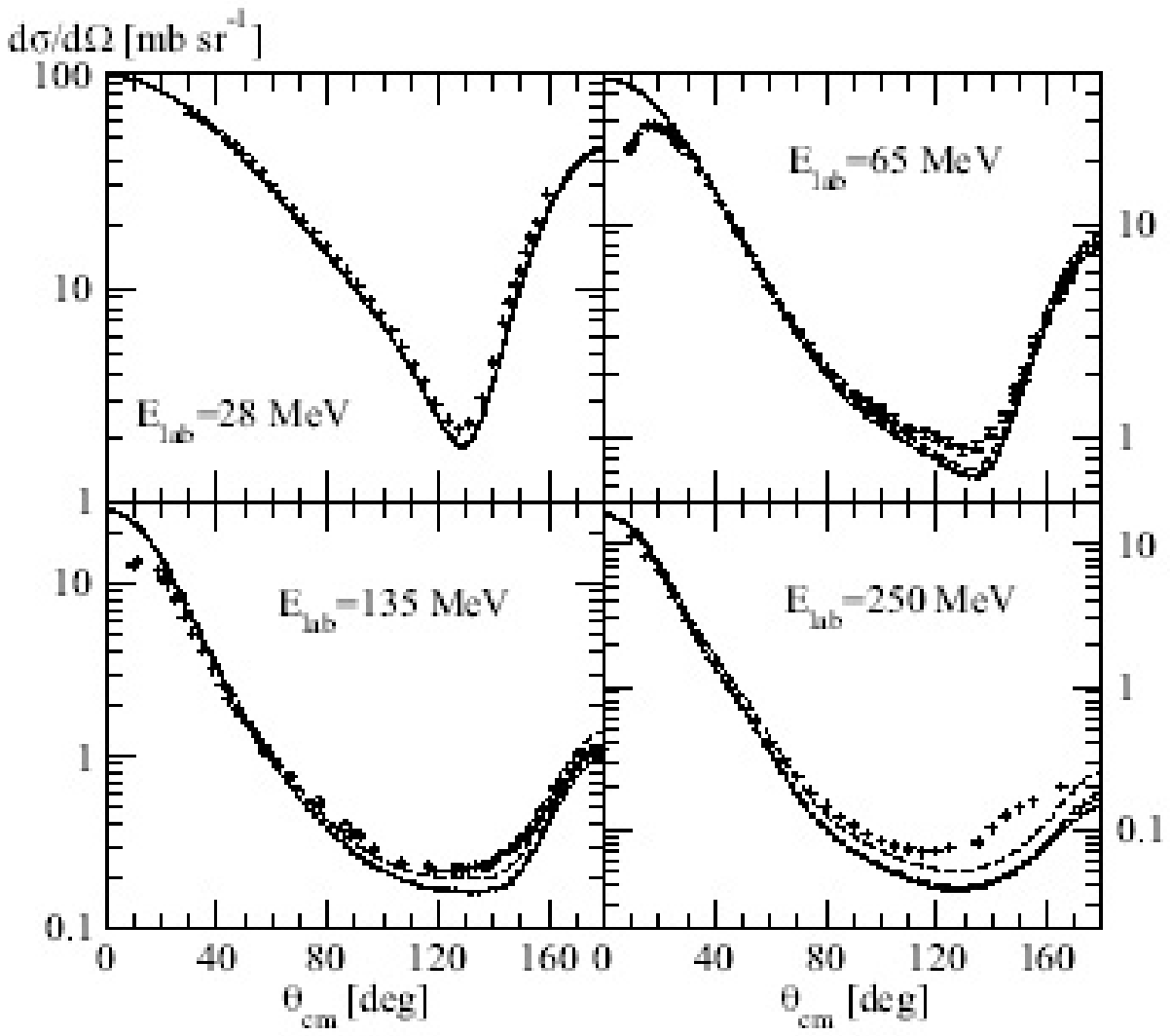}
\caption{Left: Cross section of elastic $nd$ scattering vs. c.m.
scattering angle. Short dashed line -- non-relativistic
calculation. Solid line -- non-relativistic + 3-body forces. The
figure is taken from \cite{Wit1}. Right: The same as at left for
other energies. Solid line -- non-relativistic calculation. Dashed
line -- relativistic one. The figure is taken from \cite{Wit2}.}
\label{fig3}       
\end{figure}

On the contrary, there are still the problems in the theoretical
descriptions of the three-body systems. Though the problems with
binding energy of tritium (underbinding) can be removed by
incorporating the three-body forces, there are some deviations in
description of the elastic $pd$ scattering. They are seen in fig.
\ref{fig3} taken from \cite{Wit1,Wit2}.

\begin{figure}
\centering
\includegraphics[width=5.5cm,height=5cm]{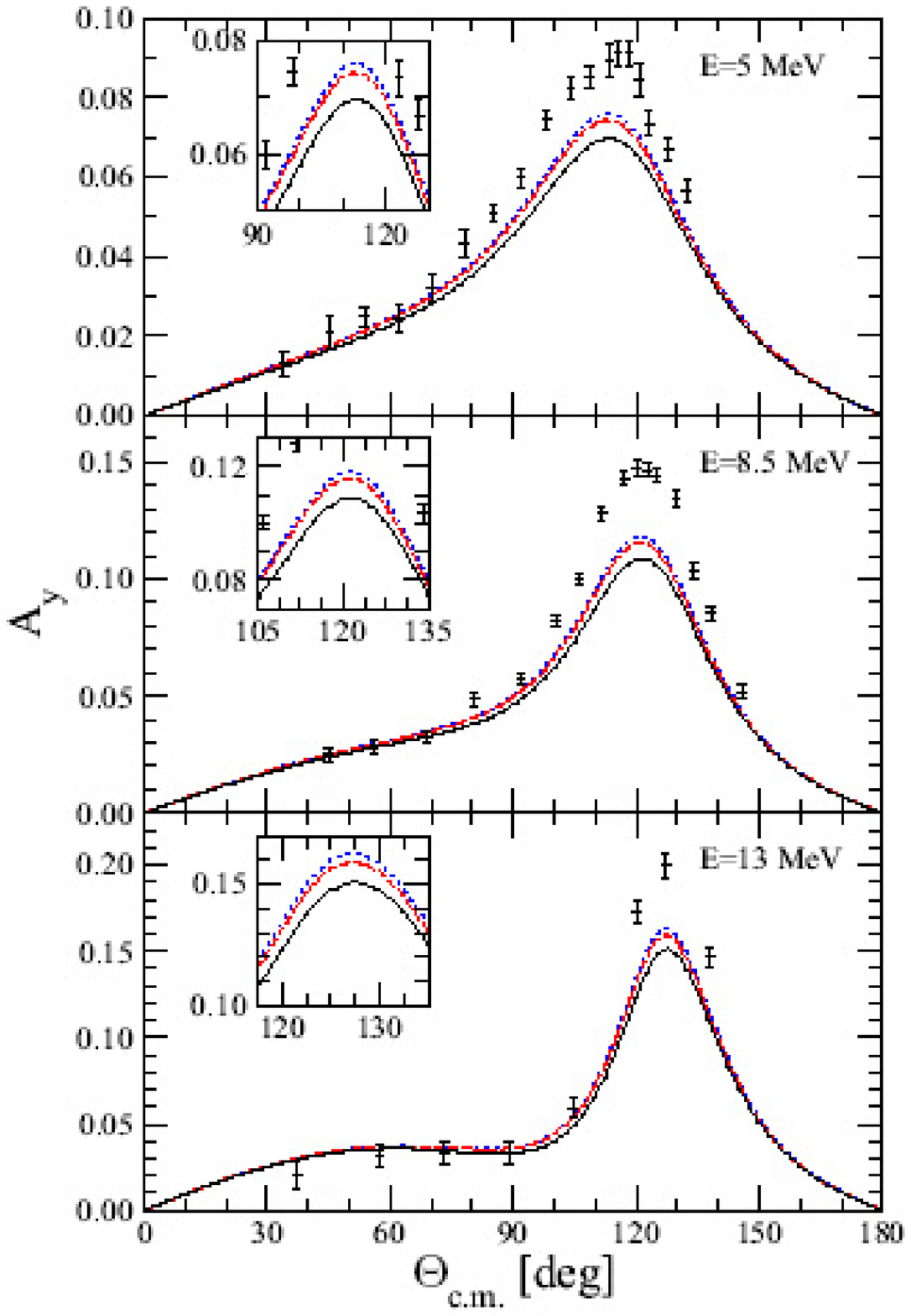}
\hspace{0.5cm}
\includegraphics[width=5.5cm,height=5.1cm]{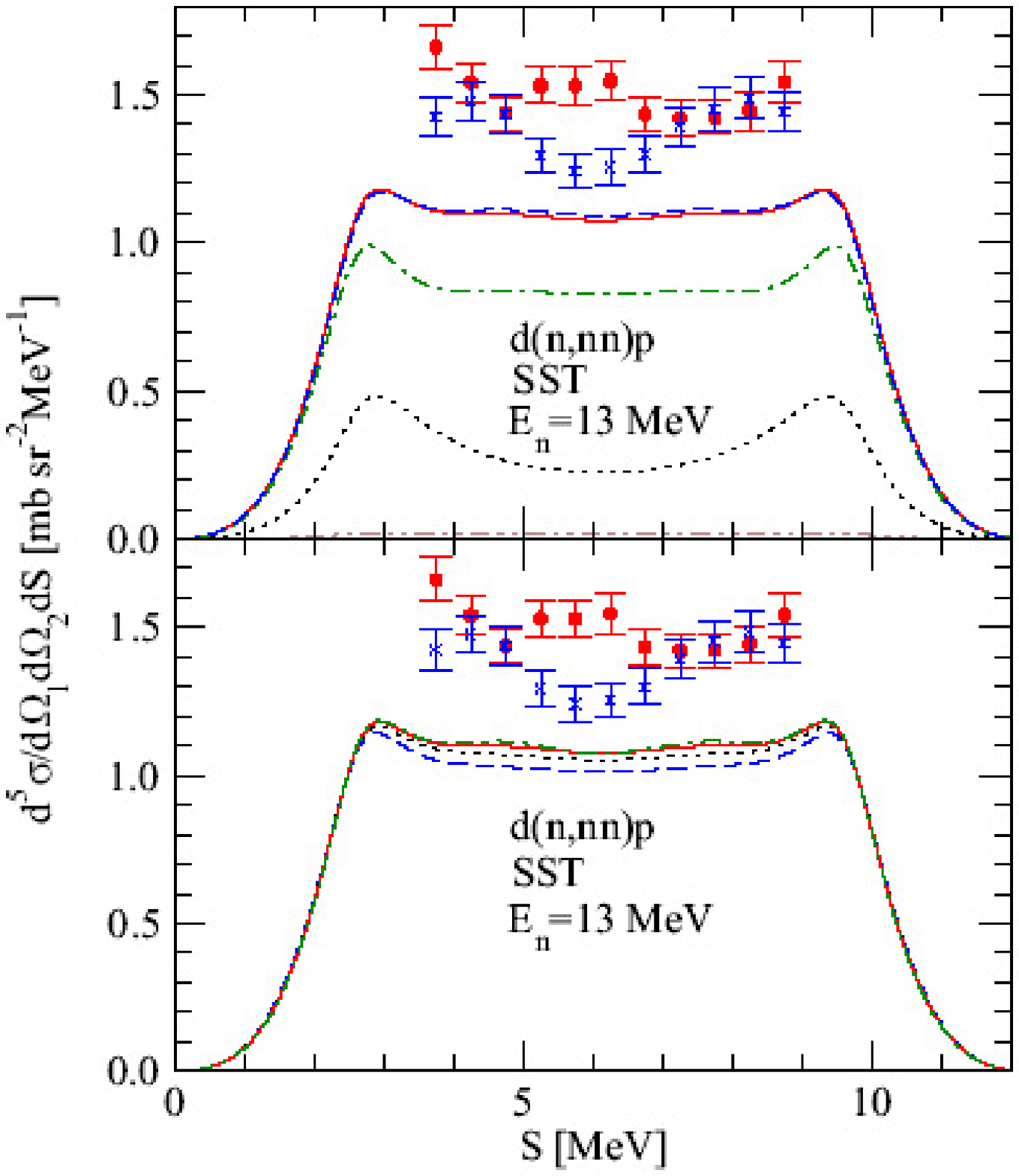}
\caption{Left: Analyzing power in elastic $pd$ scattering vs. c.m.
scattering angle. Dotted line -- non-relativistic calculation.
Solid line -- relativistic one. The figure is taken from
\cite{Wit3}. Right: The cross section of the $nd$ breakup:
$nd\to(nn)p$. The figure is taken from \cite{Wit4}.}
\label{fig4}       
\end{figure}

There is also a discrepancy in the analyzing power $A_y$ in the
$pd$ elastic scattering (see left panel in fig. \ref{fig4} taken
from \cite{Wit3}). Right panel \cite{Wit4} shows the $nd$ breakup
cross section: $nd\to(nn)p$ in a particular kinematics
corresponding to so-called symmetric space-star configuration. In
both cases, there are considerable deviations between different
versions of the theoretical calculations and experimental data.
One hopes to resolve these contradictions, properly taking into
account relativistic effects as well as three-body forces.  We
will see below that the three-body forces can be partially induced
by relativity. The importance of relativistic effects in exclusive
$pd$ breakup scattering at intermediate energies was demonstrated
in \cite{lin1,lin2}, where the relativistic Faddeev equation was
solved without employing a partial wave decomposition. The
relativistic effects improve agreement with experimental data. The
magnitude of these effects depends on configuration in the final
state. Some success in describing $A_y$ was achieved in
\cite{StGr}.

\section{Relativity in three-body systems}
The binding energy of two-body system interacting by a potential
described by the potential well $U_0$ with radius $r_0$ tends to
constant when $U_0\to\infty$, $r_0\to 0$ but $U_0r^2_0= const$. On
the contrary, for this interaction, the binding energy of
three-body system tends to $-\infty$. This is a well-known
property of non-relativistic three-body system which is called the
Thomas collapse \cite{thomas}.

It turned out that the relativity results in an effective
repulsion: for given two-body mass $M_2$, the three-body mass
$M_3$ is finite \cite{frederico,3b}. This drastic change of
three-body binding energy shows that the influence of relativity
on three-body system may be stronger than on the two-body one.

However, for enough strong interaction, corresponding to large
two-body binding energy, such that $M_2<M_c=1.43 \;m$, the
three-body mass becomes negative. In this domain of $M_2$, a
physical solution for the three-body system disappears. In the
non-relativistic scale, the binding energy equal to the total mass
of constituents, is almost infinity. Therefore the case, when
$M_3$, though being finite, approaches to zero, is a relativistic
counterpart of Thomas collapse.

\begin{figure}
\centering
\includegraphics[width=5cm,height=5cm]{resweak.eps}
\hspace{0.5cm}
\includegraphics[width=5cm,height=5cm]{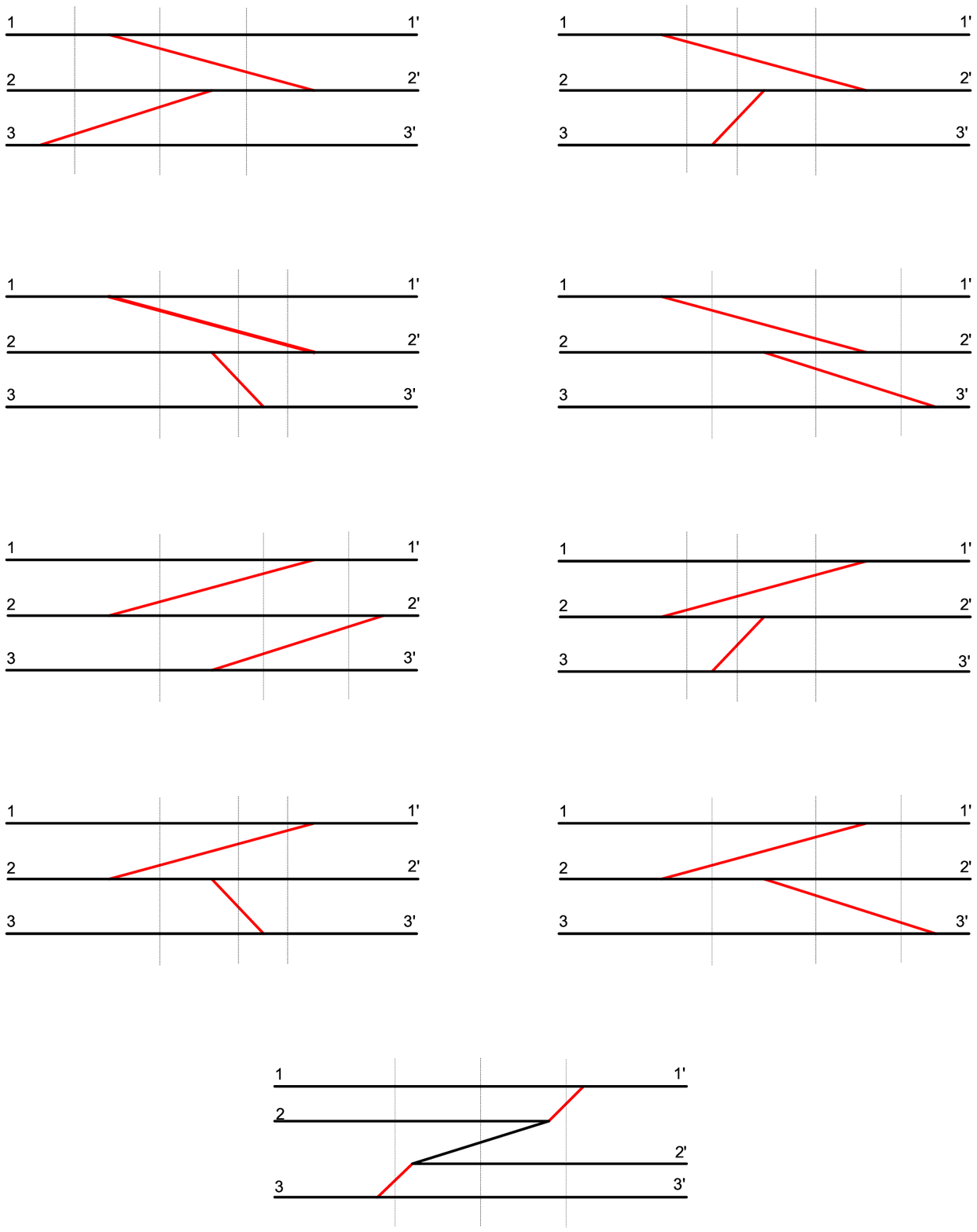}
\caption{Left: Three-body binding energy $B$ vs. coupling constant
$\alpha$, found via Shr\"odinger equation (long-dashed),
Bethe-Salpeter (solid) and the light-front (short-deshed), for
exchange masses $\mu=0.01$ and $\mu=0.5$. The results are from
\cite{maris09}. Right: The graphs contributing in the three-body
forces of relativistic origin.} \label{fig5}
\end{figure}

The relativistic three-body equations -- BS and LF ones -- have
been also solved, for spinless particles, not only for zero range
interaction, but also in more realistic case of one-boson exchange
\cite{maris09}. The corresponding two-body solution for binding
energy \cite{MC_2000} was discussed above and is shown in fig.
\ref{fig1}. The three-body binding energy vs. coupling constant
$\alpha$ is shown in fig. \ref{fig5} (left panel). In contrast to
the two-body results (see fig. \ref{fig1}), the BS and LF
calculations do not coincide, but considerably differ from each
other. They both also differ from the non-relativistic result
(like in the two-body case). However, for the two-body system,
when the exchange mass $\mu$ tends to zero, the BS and LF
calculations (which are very close to each other) tend to the
non-relativistic result. In three-body system this is not the
case.
\begin{figure}
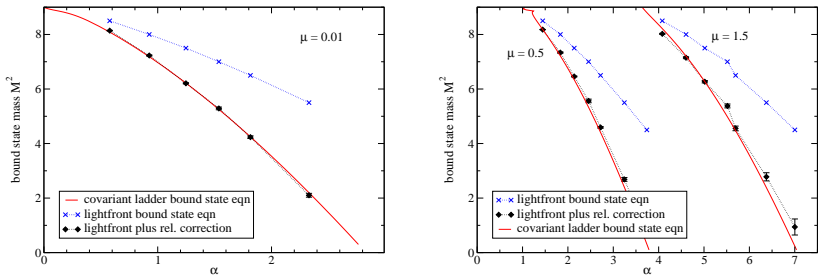

\begin{center}
\includegraphics[width=5cm]{resfinal_mu0.01.eps}
\hspace{0.5cm}
\includegraphics[width=5cm]{resfinal_mu0.5mu1.5.eps}
\caption{Three-body bound state mass squared $M_2^3$ vs. coupling
constant $\alpha$ for exchange masses $\mu = 0.01$ (left), $\mu=
0.5$ and $\mu= 1.5$ (right). The units are set by the constituent
mass: $m = 1$. The figures are taken from
\cite{maris09}.\label{fig6}}
\end{center}
\end{figure}
The reason of these deviations is  the three-body forces generated
by relativity. Corresponding graphs, containing two mesons in
flight (first considered in \cite{Yang74}), are shown in the right
panel of fig. \ref{fig5}. They are automatically included in the
three-body BS equation. However, they should be added explicitly
in the kernel of the LF equation. After taking them into account
\cite{maris09}, we find good coincidence between the BS and LF
results (see fig. \ref{fig6}). This explicitly demonstrates that
in a three-body system (a) relativistic effects and three-body
forces appear together; (b) both may be important. Notice that the
role of three-body forces may be different in different
relativistic approaches. Thus, the relativistic three-body forces
are not generated as a correction to the three-body spectator
equation \cite{Gross,stadler}. However, they should be still
incorporated as a relativistic correction to the Schr\"odinger
equation.

The relativity is not the only source of the three-body forces.
There exist other sources (e.g., the intermediate isobar creation)
which may generate "intrinsic" three-body forces. One should
include all that in the analysis of the discrepancies in
three-body reactions discussed above.

\section{Conclusions}
We conclude that relativistic effects in nuclei can be important
in spite of small binding energy. At high momenta they clearly
manifest themselves and are necessary to describe the deuteron
e.m. form factors. At the same time, there is still a discrepancy
in three-body observables which might be a result of less clarity
in understanding the corresponding relativistic effects, the
relativistic $NN$ kernel and the three-body forces.

Relativistic few-body physics remains to be a field of very
intensive and fruitful researches.




\end{document}